\documentclass[12pt]{article}
\usepackage{amsmath}
\usepackage{graphicx}
\usepackage{enumerate}
\usepackage{natbib}
\usepackage{url} 
\usepackage[colorlinks=true,linkcolor=black,anchorcolor=black,citecolor=black,filecolor=black,menucolor=black,runcolor=black,urlcolor=black]{hyperref}
\usepackage{color}

\usepackage{xcolor}
\usepackage{subcaption}
\DeclareCaptionFormat{custom}
{
    \textbf{#1#2}\textit{\small #3}
}
\newcommand{\blind}{1}

\begin{document}

\def\spacingset#1{\renewcommand{\baselinestretch}%
{#1}\small\normalsize} \spacingset{1}


\if1\blind
{
  \title{\bf Estimation of finite population  proportions for small areas -- a statistical data integration approach}
  \author{Aditi Sen and Partha Lahiri \\ University of Maryland, College Park, USA \thanks{
Aditi Sen is PhD student, Applied Mathematics \& Statistics, and Scientific Computation (e-mail: asen123@umd.edu) and Partha Lahiri is Director and Professor, The Joint Program in Survey Methodology \& Professor, Department of Mathematics, University of Maryland, College Park, MD  20742, USA (e-mail: plahiri@umd.edu).}\hspace{.2cm}}
   \maketitle
} \fi

\if0\blind
{
  \bigskip
  \bigskip
  \bigskip
  \begin{center}
    {\LARGE\bf Estimation of finite population  proportions for small areas -- a statistical data integration approach}
\end{center}
  \medskip
} \fi

\bigskip
\begin{abstract}
Empirical best prediction (EBP) is a well-known method for producing reliable proportion estimates when the primary data source provides only small or no sample from finite populations. There are potential challenges in implementing existing EBP methodology such as limited auxiliary variables in the frame (not adequate for building a reasonable working predictive model) or unable to accurately link the sample to the finite population frame due to absence of identifiers. In this paper, we propose a new data linkage approach where the finite population frame is replaced by a big probability sample, having a large set of auxiliary variables but not the outcome binary variable of interest. We fit an assumed model on the small probability sample and then impute the outcome variable for all units of the big sample to obtain standard weighted proportions. We develop a new adjusted maximum likelihood (ML) method so that the estimate of model variance doesn't fall on the boundary, which is otherwise encountered in commonly used ML method.  We also propose an estimator of the mean squared prediction error using a parametric bootstrap method and address computational issues by developing an efficient Expectation Maximization algorithm. The proposed methodology is illustrated in the context of election projection for small areas.  
\end{abstract}

\noindent%



\newpage
\spacingset{1.9} 
\section{Introduction}
\label{sec:intro}

Estimation of proportions has been a topic of great interest to statisticians for a long time. To estimate a finite population proportion from a probability sample survey, a weighted proportion is generally used, where weights assigned to the sampled units depend on the sample design.  The method is robust in the sense that modeling is not needed for obtaining estimates and associated uncertainty measures and the methodology  can be validated from the well-known randomization theory in finite population sampling; see \cite{Cochran77}.  However, when a small sample is available from the finite population, weighted proportion may yield unreasonable estimate (e.g., 0 or 1).  Moreover, in such a situation, the associated standard error (SE) estimate could be misleading (e.g., 0).

The survey-weighted proportion based on a small sample can be significantly improved using small area estimation techniques that essentially combine information from multiple sources of data using a suitable statistical model.  We refer to \cite{Rao2015}, \cite{Jiang2007}, \cite{Pfeffermann2013}, \cite{Ghosh2020} and others for a comprehensive review of small area estimation methodology and applications. 

\cite{GhoshLahiri1987} put forward an empirical Bayes (EB) approach to estimate finite population means of several strata when a simple random sample is drawn from each stratum.
Their approach covers binary data and thus can be applied to estimate several strata proportions.  
In order to implement such an EB method, one must be able to accurately identify the sampled units in the finite population frame.  Linking of the sample to the population will then allow one to partition the finite population into two segments --- sampled and unsampled units --- and predict the outcome variable of interest for the unsampled units using  the assumed model and available data. In the small area estimation context,  sampling fractions for areas are typically small and so such a linking can be avoided by simply predicting the outcome variable for all units of the finite population in a given small area; see, e.g., \cite{Rao2015}, page 304.

The finite population frame typically contains a limited auxiliary variables, which may not be adequate for building a reasonable working predictive model. Hence, we replace the finite population frame by a big probability sample that, like the frame, does not have the outcome variable of interest, 
but has the same set of auxiliary variables to fit a model that connects the outcome variable to the auxiliary variables using the small sample.
Our proposed 
empirical best prediction (EBP)
method calls for fitting an assumed 
mixed logistic
model using data from the small sample,  imputing the outcome variable for all the units of the big sample, and finally using these imputed values to obtain standard weighted proportion estimate using the big probability sample.

Parametric bootstrap method is a convenient way to estimate mean squared prediction errors (MSPE) of empirical best predictors of small area parameters; see \cite{Butar1997}, \cite{Hall2006} and \cite{ChatterjeeLahiri2013}. We develop a parametric bootstrap method for estimating MSPE of our proposed estimators with consistency properties when the number of small areas is large.

The outline of the paper is as follows. In section 2, we introduce the problem of interest and discuss point estimation. In this section, we also discuss potential working models that can be used for solving the estimation problem. In section 3, we introduce an adjusted maximum likelihood (ML) method to deal with a potential problem when model parameters are estimated on the boundary.  Our proposed adjusted ML method is indeed a nontrivial  extension of the adjusted ML methods previously developed by \cite{LahiriLi2009}, \cite{LiLahiri2010}, \cite{YoshimoriLahiri2014}, \cite{HiroseLahiri2018} and others for linear mixed models. The proposed method has a clear edge over the commonly used ML method as implemented by \cite{Bates2015glmer} in the R function \texttt{glmer} in package \texttt{lme4}. In section 4, we outline our parametric bootstrap method to estimate MSPE. In section 5, we describe the data used in the data analysis. In section 6, we carry out our data analysis.  We demonstrate the superiority  of our proposed EBP method over the direct method for predicting 2016 United States (U.S.) presidential election results for the 50 states and the District of Columbia (DC).  Finally, in section 7, we make some concluding remarks and directions for future research.

\section{PROBLEM AND METHODS}
\label{sec:methods}
Let $N_i$ be the finite population size for the $i^{th}$ area (e.g., state in a nationwide sample survey) and $m$ be the number of areas of interest. Let $Y_{ij}$ be the value of the outcome variable for the $j^{th}$ unit of the $i^{th}$ area, $i=1,\cdots,m;\;j=1,\cdots,N_i$ and our parameter of interest is $\bar Y_i=N_i^{-1}\sum_{j=1}^{N_i} Y_{ij}, \;i=1,\cdots,m.$ We have a small probability sample $\tilde s$ of size $\tilde n$ from the finite population that contains information on the outcome variable $Y$ and a vector of auxiliary variables $X$ related to $Y$ for all units. The area sample sizes $\tilde n_i$ of $\tilde s$ are small and could  be even zero for some areas. We also have a big probability sample $s$ of size $n$ from the same finite population that does not contain information on $Y$ for any unit but contains the same vector of auxiliary variables $X$. The area sample sizes $n_i$ of $s$ are large. 
For clarity, we define the two samples as:
\begin{gather*}
    \tilde{s} = \lbrace (y_{ij},x_{ij}), \;i=1, \cdots, m ;\;j=1,\cdots,\tilde{n}_i \rbrace, \\
    {s} = \lbrace x_{ij},\;i=1, \cdots, m;\; j =1, \cdots, n_i\rbrace.
\end{gather*}


Using $s$, we define the survey-weighted proportion in area $i$ as $\bar Y_{iw}=\sum_{j=1}^{n_i} w_{ij}Y_{ij},$ where $Y_{ij}$ and $w_{ij}$ are unobserved outcome variable and known survey weight (without loss of generality assuming that sum of weights is $1$ for each area), respectively, for the $j^{th}$ unit of $i^{th}$ area in the big probability sample $s\; (i=1,\cdots,m;\;j=1,\cdots,n_i).$  We assume that $n_i$ is large for each area so that  $\bar Y_i\approx \bar Y_{iw},\;i=1,\cdots,m.$ Under certain assumptions, such an approximation can be justified appealing to the law of large numbers since $n_i$’s are large for all areas $i=1,\cdots,m$. Thus, we have described how we will make use of the big probability sample survey to produce state level estimates, but as $\bar{Y}_{iw}$ is unknown we next describe how to estimate it by combining the small probability sample survey with the big probability sample survey using a working model.   

\subsection{Working model for prediction}
\label{sec:2.1}
\noindent We assume a {\it working} model for the entire finite population. The methodology developed is valid under the assumption of noninformative design as defined in \cite{sudgen1984}, \cite{pfeffermann1999} and others, so that the population working model will hold for both $s$ and $\tilde  s$. However, in this paper we use the term non-informativeness in the sense that the survey weights are no longer significant for the working model after using important auxiliary variables in the model.  This could happen, if the auxiliary variable used collectively explains most of the variations in the survey weights. The concept is similar to the one used in the paper by \cite{verret2015}.
We predict $Y_{ij}$ for all units of $s$ using information on both $Y$ and $X$ from  $\tilde s$, $X$ contained in $s$, and other area level available auxiliary variables. The working model can be fitted using $\tilde s$ because it contains information on both $Y$ and $X$ for all units. For all units in $ s$, we predict $Y_{ij}$ by $\hat Y_{ij}=E(Y_{ij}|\tilde s)$ because this will minimize the MSPE. We assume the following multi-level working model for binary outcome with normally distributed random effects for the finite population. For $i=1,\cdots,m; j=1,\cdots, N_i$,
\begin{eqnarray*}
\mbox{Level 1:}\;&&Y_{ij}|\theta_{ij} \stackrel{ind}\sim \mbox{Bernoulli} (\theta_{ij}),\\
\mbox{Level 2:}\;&&\theta_{ij}=H(x_{ij},\beta,v_i),\\
\mbox{Level 3:}\;&&v_i\stackrel{iid}\sim N(0,\sigma^2),
\end{eqnarray*}
\noindent
where $H(\cdot)$ is a known cumulative distribution function (cdf); $\beta$ is a vector of unknown fixed effects; $v_i$ is a random effect specific to the $i^{th}$ area with unknown variance component $\sigma^2$. A special case of the above is mixed logistic model in which $H(\cdot)$ is the cdf of a logistic distribution, i.e., $H(x_{ij},\beta,v_i)=\frac{\exp (x'_{ij}\beta+v_i)}{1+\exp (x'_{ij}\beta+v_i)}$.

\subsection{Empirical best prediction (EBP) approach}

Let $\gamma = (\beta,\sigma^2)$ and $\Gamma$ denote a vector of unknown parameters and  the parameter space, respectively. With known $\gamma$, the best predictor (BP) of $Y_{ij}$ for any unit in $ s$ is given by:  
$$\hat Y_{ij}^{BP}\equiv \hat Y_{ij}^{BP}(\gamma)= E(Y_{ij}|\tilde s)=E\left [E(Y_{ij}|\tilde s,\theta_{ij})|\tilde s \right ]=E\left [ H(x_{ij},\beta,v_i)\mid\tilde s) \right ],$$
where the expectation is with respect to the conditional distribution of $v_i$ given $\tilde s.$ For the assumed mixed logistic model, $\hat Y_{ij}^{BP}\equiv \hat Y_{ij}^{BP}(\gamma)=E\left [\frac{\exp (x'_{ij}\beta+v_i)}{1+\exp (x'_{ij}\beta+v_i)} \Big| \tilde s \right ].$ Let $\hat{\gamma}=(\hat\beta,\hat\sigma^2)$ be a consistent estimator of $\gamma$ for large $m$. Then EBP of $Y_{ij}$ for any unit in $s$ is given by
$\hat Y_{ij}^{EBP}= \hat Y_{ij}^{BP}(\hat{\gamma})$ and that of $\bar Y_i$ is given by $\hat{\bar Y}_{i}^{EBP}\approx \sum_{j=1}^{n_i} w_{ij}\hat Y_{ij}^{EBP}.$

\section{Adjusted maximum likelihood method}

For the general model defined in section \ref{sec:2.1}, the conditional distribution of $Y_{ij}$ given $\theta_{ij}$ is $$f(y_{ij}|\theta_{ij})=\theta_{ij}^{y_{ij}}\left(1-\theta_{ij}\right)^{1-y_{ij}}, \quad y_{ij} \in \left\{0,1\right\};  \quad i=1,\cdots,m;j=1,\cdots,\tilde{n}_i.$$
\noindent Note that 
$\theta_{ij} = H(x_{ij},\beta,v_i).$ 
Now, using independence, we obtain the conditional distribution of $Y_i=(Y_{i1},\cdots,Y_{i\tilde{n}_i})'$ given $v_i$ as
$$f(y_{i}|v_i;\beta) = \prod_{j=1}^{\tilde{n}_i}\left[
{\left\{H(x_{ij},\beta,v_i)\right\}}^{y_{ij}}{\left\{1-H(x_{ij},\beta,v_i)\right\}}^{1-y_{ij}} \right];  \quad i=1,\cdots,m.$$

\noindent The conditional distribution of $v_i$ given $y_i$ can be written as
\begin{eqnarray*}
\begin{aligned}
&f(v_i|y_i;\gamma) \propto 
f(y_i|v_i;\beta)\times f(v_i|\sigma^2) \\ 
&= \prod_{j=1}^{\tilde{n}_i} \left[ {\left\lbrace H(x_{ij},\beta,v_i) \right\rbrace }^{y_{ij}} { \left\{ 1-H(x_{ij},\beta,v_i) \right\}}^{1-y_{ij}} \times \frac{1}{\sqrt{2\pi\sigma^2}} \exp \left( {-\frac{v_i^2}{2\sigma^2}} \right)\right].
\end{aligned}
\end{eqnarray*}
\noindent For this problem $y_{ij}$ and $x_{ij}$ are observed in $\tilde s$, but the area specific random effects  $\{v_i, \;i=1 \dots m\}$ are unobserved (missing). Thus, the complete data likelihood function of $\gamma$ is 
$$L(\gamma|y,x,v) = \prod_{i=1}^{m}\prod_{j=1}^{\tilde{n}_i}\left[
{\left\{H(x_{ij},\beta,v_i)\right\}}^{y_{ij}}{\left\{1-H(x_{ij},\beta,v_i)\right\}}^{1-y_{ij}} g(v_i)\right],$$ where $g(v_i)$ is the probability density function (pdf) of $N(0,\sigma^2)$. 
\noindent But $v_i$ being unobserved, the observed data likelihood function of $\gamma$ is obtained by integrating the complete data likelihood over $v_i$'s as follows:
$$L_o(\gamma|y,x) = \prod_{i=1}^{m}\int_{-\infty}^{\infty}\left[\prod_{j=1}^{\tilde{n}_i}{\left\{H(x_{ij},\beta,v_i)\right\}}^{y_{ij}}
{\left\{1-H(x_{ij},\beta,v_i)\right\}} ^{1 - y_{ij}}g(v_i)\right]\,dv_i,\\
$$ where each integral is one-dimension, with a total of  $m$ integrals.

\subsection{Motivation for the proposed adjusted likelihood}

\noindent The maximum likelihood estimate (MLE) of $\gamma$ is obtained by maximizing $L_o(\gamma|y,x)$ with respect to $\gamma$. In our application, MLEs of $\beta$ and $\sigma^2$ yield reasonable estimates for the original sample. However, for the parametric bootstrap sample generation, we encountered two  problems while refitting the working model using the R function \texttt{glmer}. First, estimates of $\sigma^2$ for some bootstrap samples were exactly zero causing EBP to overshrink. One possible explanation is that with many good auxiliary variables, there is less variation to be captured, i.e.,  $\sigma^2$ is small. Secondly, there are cases when the algorithm for computing the MLE of $\sigma^2$ fails to converge. The problem is quite serious as in about 22\% of the bootstrap samples the algorithm for computing the MLE of $\sigma^2$ either does not converge or yields an estimate on the boundary (i.e., zero), as noted in Table \ref{tab: boundary_problem}. We thus propose a new adjusted ML method for parameter estimation, which is implemented using the Expectation Maximization (EM) algorithm applied to an adjusted likelihood function.

\begin{table}[ht]
\caption{Boundary estimation of $\sigma^2$ and converge issue when \texttt{glmer} is used in the  parametric bootstrap procedure for different bootstrap sizes. \label{tab: boundary_problem}}
\begin{center}
\begin{tabular}{rrr}
\shortstack{Bootstrap\\ Size} & \shortstack{Cases with boundary estimate\\  or convergence issue} & \% cases\\\hline
50 & 11 & 22\\
    100 & 21 & 21\\
    500 & 110 & 22\\
\end{tabular}
\end{center}
\end{table}

\subsection{Positivity of the adjusted ML estimator of $\sigma^2$}

In this subsection, for simplicity of exposition, we assume $x_{ij}=c$ and $\beta$ is known and denote $H(x_{ij},\beta,v_i)=H(v_i),\; i=1,\cdots,m; j=1,\cdots,\tilde{n}_i$. Thus the likelihood function $L(\gamma)$ reduces to $L(\sigma^2)$. We define the adjusted likelihood of $\sigma^2$ as $L_{adj}(\sigma^2)=h(\sigma^2) \times L(\sigma^2)$, where $h(\cdot)$ is a bounded smooth function such that $h(0)=0$. 
Since $H(v_i)$ is a  probability, $0 \le H(v_i)\le 1$ and $y_{ij}$ takes on values ${0,1}$ only. Thus,
\begin{eqnarray*}
\begin{aligned}
L_{adj}(\sigma^2) &= h(\sigma^2) \prod_{i=1}^{m} \int_{-\infty}^{\infty} \left[ g(v_i) \prod_{j=1}^{\tilde{n}_i} \lbrace H(v_i)  \rbrace^{y_{ij}}  \lbrace 1-H(v_i) \rbrace ^{1 - y_{ij}} \right] \,dv_i \\
&\le h(\sigma^2) \prod_{i=1}^{m} \left[ \int_{-\infty}^{\infty} g(v_i) \,dv_i\right] = h(\sigma^2).    
\end{aligned}
\end{eqnarray*}
Hence, $L_{adj}(0)=0$ when $\sigma^2=0$.  Thus, the adjusted likelihood is never maximized at $\hat{\sigma}^2 = 0$.

Next, we need to show that $\lim_{\sigma \to \infty}L_{adj}(\sigma^2)=0$. For area $i$, let $Z_i=\frac{V_i}{\sigma} \sim N(0,1)$, $\phi(z_i)$ denote the pdf of $Z_i$, and $\tilde{n}_i^*$ be the number of sampled units in $\tilde s$ for which $y_{ij}=0$. For the result to hold, we impose a condition that out of total $\tilde{n}=\sum_{i=1}^m \tilde{n}_i$ observations in $\tilde{s}$, there is at least one sampled unit for which $y_{ij}=0$, i.e., $\tilde{n}_i^* > 0$ for at least one $i$. Such a condition is not restrictive and would easily be satisfied. We define $\psi(\sigma,z_i)= H(\sigma z_i)^{\tilde{n}_i - \tilde{n}_i^*} \left(1-H(\sigma z_i)\right)^{\tilde{n}_i^*} \le 1$, as $\tilde{n}_i,\tilde{n}_i^*$ are fixed. 
Then the adjusted likelihood can be written as follows:
\begin{eqnarray*}
\begin{aligned}
L_{adj}(\sigma^2) &= 
h(\sigma^2) \prod_{i=1}^{m} \left[ \int_{-\infty}^{\infty} \phi(z_i) \prod_{j=1}^{\tilde{n}_i} \lbrace H(\sigma z_i) \rbrace^{y_{ij}} \lbrace 1 - H(\sigma z_i) \rbrace^{1-y_{ij}} \,dz_i \right]
\\
&= h(\sigma^2) \prod_{i=1}^{m} \left[ \int_{-\infty}^{\infty} \lbrace H(\sigma z_i) \rbrace ^{\tilde{n}_i - \tilde{n}_i^*} \lbrace 1-H(\sigma z_i)\rbrace ^{\tilde{n}_i^*} \phi(z_i) \,dz_i\right]
\\
&= h(\sigma^2) \prod_{i=1}^{m} \left[ \int_{-\infty}^{\infty} \psi(\sigma,z_i) \phi(z_i) \,dz_i\right].
\end{aligned}
\end{eqnarray*}

Since $h(\sigma^2)$ is bounded, we have $h(\sigma^2) \le c$ 
and  $\psi(\sigma,z) \le 1$, for any fixed z, i.e., the integral is finite. For $z>0$ and $z<0$, we make use of the functions $H(\sigma z)$ and $1-H(\sigma z)$ accordingly, which behave in opposite directions, such that limiting value is 0. For example, with logistic cdf, for $z>0$, $\lim_{\sigma \to \infty} \left\{h(\sigma^2)\times\frac{1}{1+\exp(\sigma z)}\right\} = 0$. Similarly, for $z<0$, $\lim_{\sigma \to \infty} \left\{h(\sigma^2)
\times\frac{\exp(\sigma z)}{1+\exp(\sigma z)}\right\} = 0$. 
By the  Lebesgue Dominated Convergence Theorem, we conclude for $i^{th}$ area $$\lim_{\sigma \to \infty}\int_{-\infty}^{\infty} \left[h(\sigma^2)\psi(\sigma,z_i)\phi(z_i)\right]\,dz_i = \int_{-\infty}^{\infty}\left[\lim_{\sigma \to \infty}h(\sigma^2) \psi(\sigma,z_i)\right]\phi(z_i)\,dz_i = 0.$$ Again for all other areas $i^* \neq i$, using a similar logic we can show that the limiting value is 0. Finally, using the condition on $\tilde{n}_i^*$, we get $\lim_{\sigma \to \infty}L_{adj}(\sigma^2) = 0$.  Also note that $H(\cdot)$ and $1-H(\cdot)$ are monotonic functions and $\phi(\cdot)$ being standard normal density, we have for any real numbers a and b with $-\infty<a<b<\infty$
$$L_{adj}(\sigma^2) \ge h(\sigma^2)\prod_{i=1}^{m} \left[ \lbrace H(\sigma a) \rbrace ^{\tilde{n}_i - \tilde{n}_i^*} \lbrace 1-H(\sigma b) \rbrace ^{\tilde{n}_i^*}\int_{a}^{b}\phi(z_i)\,dz_i \right] > 0.$$
Thus we have shown that the adjusted likelihood, $L_{adj}(\sigma^2)$, is 0 at $\sigma^2=0$ and tends to 0 as $\sigma^2$ goes to infinity and it has at least one positive value.

\subsection{Consistency of the adjusted MLE}
\label{sec 3.3}
Under model assumptions in section \ref{sec:2.1}, $(Y_1,\cdots,Y_m)$ are independently distributed. Hence the adjusted log-likelihood can be expressed as a sum of independent (but not identically distributed) random variables as follows:
$$
l_{adj}(\gamma|y,x) = \sum_{i=1}^{m} g(\beta,\sigma^2, Y_i) + \mbox{log} \lbrace h(\sigma^2) \rbrace = \sum_{i=1}^{m} l_i(\gamma, Y_i),
$$ 
where  $l_i(\gamma, Y_i) = g(\beta,\sigma^2, Y_i) + \frac{\log\lbrace h(\sigma^2) \rbrace }{m}$ and
$$  
g(\beta,\sigma^2, Y_i) = \mbox{log} \left ( \int_{-\infty}^{\infty}\left[\prod_{j=1}^{\tilde{n}_i}{\left\{H(x_{ij},\beta,v_i)\right\}}^{y_{ij}}{\left\{1 - H(x_{ij},\beta,v_i)\right\}} ^{1 - y_{ij}}\frac{1}{\sqrt{2\pi\sigma^2}} \exp \left( {-\frac{v_i^2}{2\sigma^2}} \right)\right]\,dv_i\\ \right ).$$
 Let $\gamma_0$ be the true value and $\gamma_u$ be the $u^{th}$ component of $\gamma$. We assume the following regularity conditions: 
\begin{enumerate}
    \item [(i)] $\gamma_0$ is an interior point of $\Gamma$; 
    \item [(ii)] Each $l_i$ is three times continuously differentiable with respect to $\gamma$ in some neighborhood of $\gamma_0$;
    \item [(iii)] $E \left (\frac{\partial l_i}{\partial \gamma}|_{\gamma_0} \right ) = 0; \quad  1 \le i \le m$;
    \item [(iv)] $E \lbrace {(\partial l_i/ \partial \gamma_u)|_{\gamma_0} \rbrace }^2$, $E{ \lbrace ( \partial^2 l_i / \partial \gamma_u \partial \gamma_v ) |_{\gamma_0} \rbrace }^2$ , $ E \sup_{|\gamma - \gamma_0| \le \epsilon} |\partial^3 l_i / \partial \gamma_u \partial \gamma_v \partial \gamma_w  |, \quad 1 \le i \le m, \quad 1 \le u,v,w \le r $,  are bounded; where $\epsilon$ is some positive number and $r$ is the dimension of $\gamma$; and
    \item [(v)] $\frac{1}{m} \sum_{i=1}^{m} E (  \frac{\partial^2 l_i}{\partial \gamma^2} |_{\gamma_0}) \to -B $ , where B is positive definite.
\end{enumerate}
Then with probability tending to one, there is a solution $\hat{\gamma}$ (adjusted MLE) to $\frac{\partial l_{adj}} {\partial \gamma} = 0$  in $\Gamma$ such that $\hat\gamma \overset{P}{\to} \gamma_0$  as $m \to \infty$. Since Bernoulli distribution belongs to exponential family, the above follows from Theorem 1 in \cite{Jiang2001}.

\subsection{EM algorithm and its application to this problem}
\label{sec : 4.1.1}
After solving the boundary problem, we can  use Monte-Carlo approximation of the integration by a summation in the adjusted likelihood function. First, we need to predict $v$. Here we apply the idea of the EM algorithm (refer to \cite{LittleRubin2015}) where we treat $v$ as missing and $(y,x)$ as observed.

Let $l(\gamma|y)$ be the log-likelihood when there are no missing values. Let $y_{(0)}$ be observed data and $y_{(1)}$ be unobserved data so that the complete data is $y=(y_{(0)},y_{(1)})$. We want to maximize $l(\gamma|y)$ over $\gamma$. If $\gamma^{(t)}$ is the estimate of $\gamma$ at iteration $t$ then E and M steps of the EM algorithm are as follows:

\noindent \textbf{E Step:} Obtain the expected value of complete likelihood $l(\gamma|y)$ under the distribution of $(y_{(1)}|y_{(0)})$, which is a function of $\gamma$ given $\gamma^{(t)}$ as follows:
$$\\
Q(\gamma|\gamma^{(t)})=E_{\gamma^{(t)}}\left [l(\gamma|y)|y_{(0)}\right ]=\int l(\gamma|y)f_{\gamma^{(t)}}(y_{(1)}|y_{(0)})dy_{(1)}.$$

\noindent \textbf{M Step:} Maximize the above function $Q(\gamma|\gamma^{(t)})$ of $\gamma$ to obtain the estimate $\gamma^{(t+1)}$ as $Q(\gamma^{(t+1)}|\gamma^{(t)})=\mbox{max}_{\gamma} Q(\gamma|\gamma^{(t)})$.
\noindent In this problem we want to estimate $\gamma$ with given or known values as $y_{(0)} = (y,x)$ and missing or unknown values $y_{(1)} = v$.
Now we align this problem in the framework of EM algorithm. 

\noindent \textbf{E step}: 
\begin{enumerate}
    \item [Step 1.] Assuming some starting values of $\gamma$, say $\gamma^{(0)}$, in the E-step we need to calculate $Q(\gamma|\gamma^{(0)})$. For the $i^{th}$ area, due to missing values of $v_i$ we first approximate the distributional form of $(v_i|y_i)$ (using Laplace technique) by a Normal distribution with mean $\hat{v_i}=\arg\max_{v_i}f({v_i}|y_i)$ 
    and variance given by $\hat{\tau}_i^{2}={ \left[-\frac{\partial^2 \mbox{log} f({v_i}|y_i;\gamma)} {\partial v_i^2}|_{\hat{v_i}}\right]}^{-1},$ which is the inverse of the observed information i.e.
    \begin{eqnarray*}
    \begin{aligned}
        f(v_i|y_i,\gamma) 
        &\propto \exp \left(-\frac{v_i^2}{2\sigma^2} \right) \prod_{j=1}^{\tilde{n}_i} \left[ 
        {\lbrace H(x_{ij},\beta,v_i) \rbrace }^{y_{ij}}
        {\lbrace 1-H(x_{ij},\beta,v_i)\rbrace}^{1 - y_{ij}}
        \right ] \\
        & = k(v_i)\approx N\left( \hat{v_i},\hat{\tau}_i^{2} \right).   
    \end{aligned}
    \end{eqnarray*}
        
    \item [Step 2.] To calculate $\hat{v_i}$ for $i=1, \cdots, m$, we numerically maximize the function $k(v_i)$ of $v_i$ and hence obtain $\hat{v_i}$ where the maximum value of $k(v_i)$ is achieved. To calculate $\hat{\tau}_i^{2}$, we first differentiate $-\mbox{log}\left( f(v_i|\gamma^{(0)})\right)$ or equivalently $-\mbox{log}\left(k(v_i|\gamma^{(0)})\right)$ twice with respect to $v_i$, then plug in values of $\hat{v_i}$ and take inverse. 
    We know 
    \begin{eqnarray*}
    \mbox{log }k({v_i})=-\frac{v_i^2}{2\sigma^2}+\sum_{j=1}^{\tilde{n}_i}\left[ y_{ij} \mbox{log} \left\{H(x_{ij},\beta,v_i)\right\} + \left(1 - y_{ij}\right) \mbox{log} \left\{1-H(x_{ij},\beta,v_i)\right\}\right]
    \end{eqnarray*}
    
    For $H(x_{ij},\beta,v_i)$ as logistic cdf, the derivatives are as follows:\\
    $\frac{\partial \mbox{log} k({v_i})} {\partial v_i} = -\frac{v_i}{\sigma^2}+\sum_{j=1}^{\tilde{n}_i}\left[\frac{y_{ij}}{1+\exp(x'_{ij}\beta + v_i)} - (1-y_{ij}) \left\{\frac{\exp(x'_{ij}\beta + v_i)}{1+\exp(x'_{ij}\beta + v_i)}\right\} \right]\\
    \frac{\partial^2 \mbox{log} k({v_i})} {\partial v_i^2} = - \frac{1}{\sigma^2} - \sum_{j=1}^{\tilde{n}_i}\frac{\exp(x'_{ij}\beta + v_i)}{\left\{1+\exp(x'_{ij}\beta + v_i)\right\}^2} \Rightarrow 
    \hat{\tau_i}^2=\left[\frac{1}{\sigma^2} + \sum_{j=1}^{\tilde{n}_i}\frac{\exp(x'_{ij}\beta + \hat{v_i})}{\left\{1+\exp(x'_{ij}\beta + \hat{v_i})\right\}^2}\right]^{-1}$

    \item [Step 3.] Draw sample $z_i,\;i=1,\cdots,m$, independently $R$ times from $N(0,1)$ and obtain sample $\tilde{v_i}$ from $N(\hat{v_i},\hat{\tau}^2_i)$ using  the formula $\tilde{v_i}=\hat{v_i}+\hat{\tau}_i\times z_i,\; i=1,\cdots,m$. 
    \item [Step 4.] We approximate $E_{\gamma^{(0)}}\left [l(\gamma|y,v)\right ].$ Having obtained values of $\tilde{v_i}$, i.e., samples from the distribution $f(v_i|\gamma^{(0)})$, we can approximate the expected complete data log-likelihood with respect to the distribution of $f(v_i|\gamma^{(0)})$. 
    Here we use Monte-Carlo approximation methods and also adjust the log-likelihood to avoid the problem of boundary estimates. Expressing log-likelihood as sum of separate functions of $\sigma^2$ and $\beta$ for ease of computation, we obtain the function to maximize in the M step as follows:
    \begin{eqnarray*}
    \begin{aligned}
    Q(\gamma|\gamma^{(t)}) 
    &= E_{\gamma^{(t)}} \left[ l(\gamma|y)|y_{(0)} \right] = E_{\gamma^{(t)}} [l(\gamma|y,x,\tilde{v})] \\ &\approx
    \frac{1}{R} \sum_{r=1}^{R} l_1(\sigma^2|\tilde{v}_{r}) + \frac{1}{R} \sum_{r=1}^{R} l_2(\beta|y,x, \tilde{v}_{r}),
    \end{aligned}
    \end{eqnarray*}
    where
    $l_1(\sigma^2|v_{r}) =  \mbox{log}\left\{h(\sigma^2)\right\}-\frac{\sum_{i=1}^{m}\tilde{n}_i}{2}\log \left(\sigma^2\right) - \sum_{i=1}^{m} \frac{\tilde{n}_i\tilde{v}_{ri}^2}{2\sigma^2} \mbox{ and} \\
    l_2(\beta|y,x,\tilde{v}_{r}) = \sum_{i=1}^{m}\sum_{j=1}^{\tilde{n}_i}\left[{  y_{ij} \mbox{log}\left\{H(x_{ij},\beta, \tilde{v}_{ri}) \right\}
    + \left(1 - y_{ij}\right) \mbox{log} \left\{1-H(x_{ij},\beta,\tilde{v}_{ri})\right\}}\right].$
    \end{enumerate}

\noindent \textbf{M step}: Now we have a function of $\gamma$ that we need to maximize to obtain our next iterate $\gamma^{(1)}$. We can do this in R using optimization functions available in existing packages.   

\section{MSPE USING PARAMETRIC BOOTSTRAP}
\label{sec: par boot}
We now discuss estimation of MSPE of EBP using a parametric bootstrap method. Let $\{\tilde s_b; b=1,\cdots,B\}$ denote parametric bootstrap resamples from the small sample $\tilde{s}$. To generate these $\tilde s_b$ from $\tilde s$ we first independently generate $v_{bi}$ from $N(0,\hat\sigma^2)$, then  compute $\theta_{bij}=H(x_{ij},\hat\beta,v_{bi})$ and finally, we independently generate $y_{bij}$ from Bernoulli$(\theta_{bij})$,\\$b=1,\cdots,B;\; i=1,\cdots,m;\;j=1,\cdots,\tilde n_i.$  Using the $b^{th}$ parametric bootstrap sample $\tilde s_b$, we can re-estimate $\gamma$ by $\hat{\gamma}_b=(\hat\beta_{b},\hat\sigma^2_{b})$, and eventually produce EBP of $Y_{ij}$, denoted by $\hat Y_{bij}^{EBP}$ based on $\tilde s_b$, auxiliary variables in $s$, and area level auxiliary variables from other sources.  Thus EBP of $\bar Y_i$ based on the $b^{th}$ bootstrap sample $\tilde s_b$ is given by $\hat{\bar Y}_{bi}^{EBP}\approx \sum_{j=1}^{n_i} w_{ij}\hat Y_{bij}^{EBP}.$ For  $s$, we can also generate $y_{bij},\;b=1,\cdots,B; i=1,\cdots,m; j=1,\cdots,n_i$ and define $\bar y_{biw}\approx \sum_{j=1}^{n_i} w_{ij}y_{bij}.$ We propose the following parametric bootstrap estimator of MSPE:
$$ \widehat{\mbox{MSPE}}(\hat{\bar Y}_{i}^{EBP} )=\frac{1}{B} \sum_{b=1}^B \left [ \hat{\bar Y}_{bi}^{EBP} -  \bar y_{biw}   \right ]^2,$$ where $\hat{\bar{Y}}_i^{EBP}$ is EBP of $\bar{Y}_i$, the true proportion of the $i^{th}$ area.
Square root of $\widehat{\mbox{MSPE}}(\hat{\bar Y}_{i}^{EBP} )$ can be treated as the estimated SE of EBP of $\bar Y_i$.

\subsection{Consistency of the parametric bootstrap method}

We can reasonably expect MSPE of EBP to be a function of model parameters $\gamma$, i.e., $ MSPE(\hat{\bar{Y}}_i^{EBP}) = E(\hat{\bar{Y}}_i^{EBP}-\bar{Y}_i)^2 = q_i(\gamma)
$,
where the expectation is with respect to the assumed model and $q_i$ is a smooth function.
Let $\hat{\bar{Y}}_{bi}^{EBP}$ be the EBP of $\bar{Y}_i$ from the $b^{th}$ bootstrap sample. Following the Weak Law of Large Numbers, we have, for large $B$,
$$
\frac{1}{B}\sum_{b=1}^{B}(\hat{\bar{Y}}_{bi}^{EBP}-\bar y_{biw} )^2 \overset{P} {\to} E_\ast(\hat{\bar{Y}}_i^{EBP\ast}-\bar{Y}_i^{\ast})^2=q_i(\hat\gamma),
$$
where the convergence is in probability, the expectation $E_*$ is with respect to the parametric bootstrap distribution, and $\hat{\gamma}$ is adjusted MLE of $\gamma$. Using Taylor series expansion and ignoring higher order terms, we obtain 
$q_i(\hat\gamma)-q_i(\gamma) \approx q'_i(\gamma^*)(\hat\gamma-\gamma) \overset{P}{\to} 0$  as $m \to \infty$,   where $\gamma^*$ is between $\hat{\gamma}$ and $\gamma$ and the convergence is with respect to the assumed model. Because $\hat\gamma \overset{P}{\to} \gamma$  as $m \to \infty$, i.e., the adjusted ML estimators are consistent as shown in section \ref{sec 3.3}.  Thus conclude $$E_\ast(\hat{\bar{Y}}_i^{EBP\ast}-\bar{Y}_i^{\ast})^2 \overset{P}{\to} E(\hat{\bar{Y}}_i^{EBP}-\bar{Y}_i)^2,\;\mbox{as} \;m \to \infty$$
where the convergence in probability is with respect to the parametric bootstrap model. For more information, we refer to \cite{Lahiri2023}, where the authors discuss estimation of a general class of uncertainty measures for a small area estimator (not necessarily an EBP). 

\section{DATA}

\begin{figure}[ht]
    \centering
    \caption{Multiple survey data structure in comparison to the finite population.}
    \label{fig: data_structure}
    \includegraphics[width=0.7\linewidth]{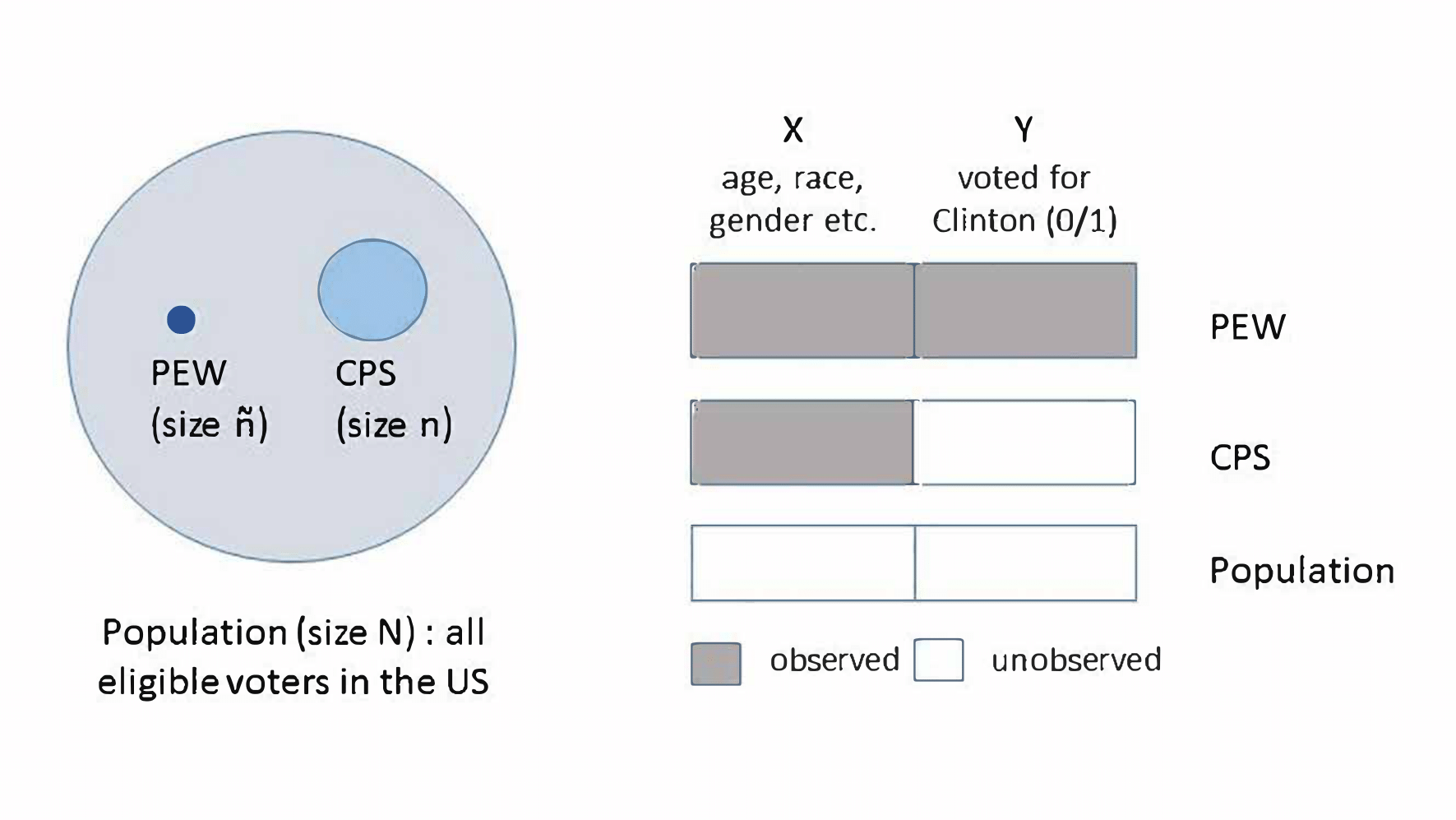}
\end{figure}

\noindent We illustrate the methodologies, discussed in preceding sections, in predicting percentage of voters for Clinton in the 50 states and DC in the 2016 U.S. Presidential election. Our main data sources are two survey data. The first is the Pew Research Organization's data, containing information about the outcome variable (voting preference) and the second being the Current Population Survey (CPS) data. We also have the actual results from the 2016 election, which helps us comparing the performances of different estimators. Figure \ref{fig: data_structure} offers a possible situation involving two independent surveys and the finite population. It also signifies the presence and absence of outcome and auxiliary variables in different databases.

\subsection{Pew and CPS Survey data}

The Pew Research Center provides information on social issues, public opinion, and demographic trends shaping the U.S. and the world. We consider their October 2016 Political Survey conducted during October 20-25, before the 2016 Presidential election. It is a telephone survey of national samples with a combination of landline and cellular random digit dial (RDD) samples containing 2,583 interviews.  We obtain this data for 1.6k registered voters from $49$ areas ($48$ states and DC); there is no sample for the states of Montana (MT) and South Dakota (SD). The Pew survey contains information on voting preference along with individual characteristics like age, gender, race, etc. This is a probability sample where the weights are obtained from a two-stage weighting procedure involving calibration. For more information on sample design, weighting methodology, etc. refer to the methodology document on the \href{https://www.people-press.org/dataset/october-2016-political-survey/}{Pew website}. We create a binary outcome variable taking the value 1 if a person prefers to vote for Clinton and 0 otherwise. We want to estimate the percentage of voters for Clinton in the U.S. states and DC. Direct survey  estimates would not serve the purpose well as such estimates may not be trustworthy due to high standard errors (SE). Moreover, direct estimates are nonexistent for two states as there is no observation from the Pew sample for these two states. So we take the help of a big survey.

The Current Population Survey (CPS) is jointly sponsored by the U.S. Census Bureau and the U.S. Bureau of Labor Statistics (BLS). In addition to being the primary source of monthly labor force statistics, the CPS collects data for a variety of other studies by adding a set of supplemental questions to the monthly basic questions. The target population consists of all persons in the civilian noninstitutional population of the United States living in households, 16 years of age and older. Supplemental inquiries are conducted annually or biannually and cover a wide range of topics such as child support, volunteerism, health insurance coverage and school enrollment. The November 2016 Voting and Registration Supplement of CPS is used for this analysis, where in addition to the basic CPS questions, interviewers asked supplementary questions on voting and registration to all persons 18 years of age and older. It contains data of about 58.5k individuals from 50 states and DC. The full dataset can be found at the \href{https://www.nber.org/cps/}{National Bureau of Economic Research website}. For further details on the survey according to \cite{pricssa}, the Preferred Reporting Items for Complex Sample Survey Analysis (PRICSSA) checklist including sample design, response rate, survey questions, estimation/weighting, variance estimation methodology etc. refer to the supplemental material.

We obtain auxiliary variables for the analysis from both Pew and CPS data. We consider individual demographic characteristics like age, gender, race, etc., which are converted into categorical variables. We also consider a state level auxiliary variable: the state wise percentage of voters for Obama in the 2012 Presidential election. Only the Pew sample contains information on the outcome variable, so we can view this as a missing data problem as well as prediction problem because CPS does not contain information on voting preference. We use Pew data for fitting the assumed model and use CPS for predicting the outcome at state level. 

\subsection{Actual 2016 election result}
In this problem we can actually evaluate the prediction from our proposed method and compare such estimates with true values as we already have the 2016 Presidential election results available \href{https://www.uselectionatlas.org/}{here}. We consider the actual percentage of voters for Clinton for 50 states and DC and observe from the data that the lowest value of the outcome variable is in Wyoming (WY) with 22\% and the highest is in DC with 91\%.

\section{DATA ANALYSIS}

This section is divided into three subsections: where we compute the area level point estimates using proposed algorithm, then compare with direct estimates and actual values and finally calculate MSPE using parametric bootstrap. The codes for all sections of the data analysis are available at this \href{https://github.com/asen123/data_integration}{website}. 

\subsection{Point estimation using new algorithm}
\label{sec: EM func}

For the working model we consider auxiliary variables described in Table \ref{tab: cov_list}. We did preliminary analysis using the \texttt{glmer} function in R to fit a mixed logistic model with all available auxiliary variables and states as random effects. Herein we observe that among the auxiliary variables, the age group 45-64 years, gender, race, education college (graduate, postgraduate) and voting percentage for Obama are significant at 5\% level of significance, so we consider only these auxiliary variables in the final working model. Further details on parameter estimates, standard errors, p-values, goodness-of-fit statistics for the model are provided in supplemental material.

\begin{table}[ht]
\caption{List of unit level auxiliary variables with detailed levels. \label{tab: cov_list}}
\begin{center}
\begin{tabular}{rrr}
Auxiliary variable & No. of Levels & Levels\\\hline
Age (years) & 4 & 18-29, 30-44, 45-64 and 65+\\
     Gender & 2 & Male or Female\\
     Race & 3 & White, Black or Hispanic\\
     Education & 4 & \shortstack{Higher Secondary, Some college, \\ College Grad or Postgrad}\\
     Region & 4 & Northeast, South, North Central or West\\
\end{tabular}
\end{center}
\end{table}

We write a user-defined function in R, which takes a dataset and initial values of model parameters $\gamma$, as inputs and provides the estimates $\hat{\gamma}$, as outputs. We use this function iteratively, i.e., in step $t$ the function outputs $\hat{\gamma}^{(t)} = (\hat{\beta}^{(t)}, \hat{\sigma}^{2^{(t)}})$, which will be used as initial values for step $(t+1)$, wherein the function will output $\hat{\gamma}^{(t+1)} = (\hat{\beta}^{(t+1)}, \hat{\sigma}^{2^{(t+1)}})$ and so on, until convergence. Convergence is attained when the absolute difference between the parameter estimates in subsequent steps is less than 0.01 i.e. $|\hat{\sigma}^{2(t+1)}-\hat{\sigma}^{2(t)}| < 0.01 \mbox{ and } |\hat{\beta}_k^{(t+1)}-\hat{\beta}_k^{(t)}| < 0.01, \; k=0,\cdots,7$. We elaborate the steps of the algorithm below:
\begin{itemize}

\item [Step 1.] First we use initial set of values for $(\beta_0,\cdots,\beta_7,\sigma^2)$ and compute $x'\beta$ which will be required in the next step in the functional form of $f(v)$ as defined in section \ref{sec : 4.1.1}.

\item [Step 2.] For small area $i$, we maximize $f(v_i)$ using R function \texttt{optimize} and obtain $(\hat{v}_i,\hat{\tau}_i^2), i=1,\cdots, m$. \texttt{Optimize} is a useful function where the optimization method used is a combination of golden section search and successive parabolic interpolation. But the method works in the one variable optimization problem only. We have used bounds in the \texttt{optimize} function in order to avoid unrealistic values of $\hat{v}$. 

\item [Step 3.] We draw for $i^{th}$ area an observation from $N(\hat{v}_i,\hat{\tau}_i^2)$ and repeat this $R=100$ times to get $(\tilde{v}_{1i},\cdots,\tilde{v}_{100i})$ for each area. The same thing is done for $m=51$ areas.

\item [Step 4.] Since we can separate $Q(\gamma|\gamma^{(t)})$ into functions of $\beta$ and $\sigma^2$ as described in section \ref{sec : 4.1.1}, we maximize separately. Note that this is based on the adjusted likelihood, and we have used $h(\sigma)=\sigma^2$ with bounds to satisfy the conditions. We plug in the above values and maximize $Q(\sigma^2|\sigma^{2^{(t)}})$, again using R function \texttt{optimize}, which outputs $\hat{\sigma}^{2{(t)}}$ for the $t^{th}$ iteration. 

\item [Step 5.] We maximize $Q(\beta|\beta^{(t)})$, which is an objective function in a multidimensional space and we use R function \texttt{optim}, which is a general-purpose optimization function based on various methods for solving unconstrained nonlinear optimization problems. We tested both the Nelder-Mead and the Broyden-Fltecher-Goldfarb-Shanno (BFGS) methods, of which the latter is better in terms of quicker computation performance. Our benchmark is to compare the process with existing R function of fitting mixed logistic model, \texttt{glmer}, which takes a few seconds to run. So we try various parallel optimization packages and find \texttt{optimParallel}, which is a parallel version of the L-BFGS-B method of \texttt{optim}, to be particularly useful (refer to \cite{GerberOptimParallel}). Using this the time to converge for the EM algorithm reduces to 5 minutes.
\end{itemize}

We first test the algorithm through a simulation study where we generate data from known distributions and check the output estimates $\hat{\gamma}$ from our EM function and also compare the values with \texttt{glmer} outputs, which are our benchmarks. As described in the mixed logistic model set up in section \ref{sec:2.1}, we first generate $v_i$ from $N(0,\sigma^2)$, $x_{ij}$ from standard normal distribution, compute $\theta_{ij}=\frac{\exp (x'_{ij}\beta+v_i)}{1+\exp (x'_{ij}\beta+v_i)}$, then generate $y_{ij}$ from $Bern(\theta_{ij})$. We check this for different values of $m$ (number of small areas), e.g. $m=20,50,75 \mbox{ and } 100$ and different initial values of $\gamma$. In all these cases, we find that the algorithm works well both in terms of number of iterations and time taken to converge as well as estimated value of parameters. 

\begin{figure}[ht]
    \centering
    \caption{Model parameter estimates $(\hat{\sigma},\hat{\beta_0},\hat{\beta_1})$ vs iteration where prefix EM denotes the proposed EM algorithm and prefix GLMER denotes existing method.}
    \label{fig: EM_plot}
    \includegraphics[width=0.8\linewidth]{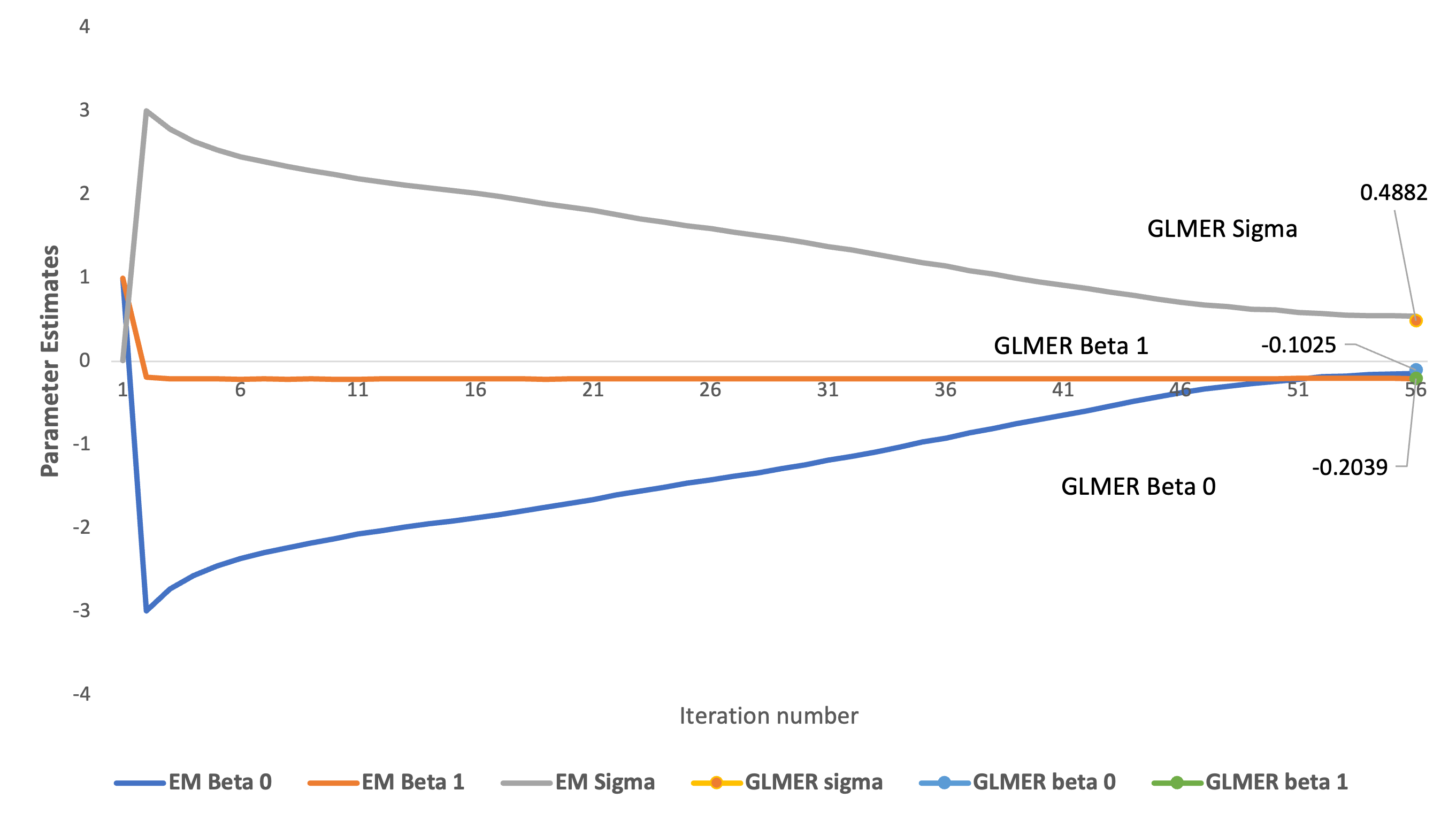}
\end{figure}

To finalize the algorithm, for each of the different optimization methods, we look at the estimated parameter values and compare with the \texttt{glmer} outputs on Pew data.  The algorithm produces comparable estimates using \texttt{optim} and best time is achieved using BFGS method. Ultimately, we prefer the algorithm using parallel computing, i.e. \texttt{optimParallel}, as time to converge reduces significantly and estimates are similar to \texttt{glmer}. We provide the details of such findings in Tables \ref{tab: optimization details} and \ref{tab: est val}. 
Figure \ref{fig: EM_plot} shows the convergence of the algorithm using one auxilliary variable on Pew data. The EM algorithm takes 55 iterations to converge, with initial values given by $\sigma=0.01$, $\beta_k = 0.1,\;k=0,1$. Computation time and total iterations to convergence depend on the initial values of parameters.

\begin{table}[ht]
\caption{Algorithm details for 4 methods M1, M2, M3, M4 where only M1 uses ML technique for model parameter estimation wheres M2, M3 and M4 uses adjusted ML with different optimization techniques. \label{tab: optimization details}}
\begin{center}
\begin{minipage}{\textwidth}
\begin{tabular}{rrrrr}
\shortstack{Model parameter\\ estimation techniques} & Method & \shortstack{R function \\  \& package} & \shortstack{Optimization \\ techniques} & \shortstack{Average time \\ to converge}\\ 
\hline \hline
\shortstack{ML}\footnote{Maximum likelihood method for generalized linear mixed model (GLMM)} & M1 & \texttt{glmer(lme4)} & \shortstack{adaptive \\ Gauss-Hermite \\quadrature} & \shortstack{less than \\ 1 min}\\
\hline
\shortstack{Adjusted ML}\footnote{Proposed new method using EM algorithm for GLMM} & M2 & \texttt{optim} & Nelder Mead & 8 hrs\\
 & M3 & \texttt{optim} & BFGS & 2 hrs\\
 & M4 & \texttt{optimParallel}  & L-BFGS-B & 5 mins
\end{tabular}
\end{minipage}
\end{center}
\end{table}

\begin{table}[ht]
\caption{Parameter estimates from existing method using R function \texttt{glmer} denoted as M1 and proposed method using EM algorithm and \texttt{optimParallel} denoted as M4. \label{tab: est val}}
\begin{center}
\begin{tabular}{rrrrrrrrrr}
Method & $\hat{\sigma}$ & $\hat{\beta}_0$ & $\hat{\beta}_1$ & $\hat{\beta}_2$ & $\hat{\beta}_3$ & $\hat{\beta}_4$ & $\hat{\beta}_5$ & $\hat{\beta}_6$ & $\hat{\beta}_7$\\
\hline
M1 & 0.19 & -0.96 & -0.22 & 0.63 & 3.01 & 1.13 & 0.42 & 0.99 & 1.10\\
M4 & 0.13 & -0.96 & -0.21 & 0.61 & 2.98 & 1.16 & 0.42 & 0.98 & 1.12\\
\end{tabular}
\end{center}
\end{table}

\subsection{Comparison with direct estimates and actual values}

\begin{figure}
    \centering
    \caption{Actual and predicted percentages of voters for Clinton in the 2016 U.S. Presidential Election from direct and EBP method in 50 states and DC.}
    \label{fig: all_state}
    \includegraphics[width=0.7\linewidth]{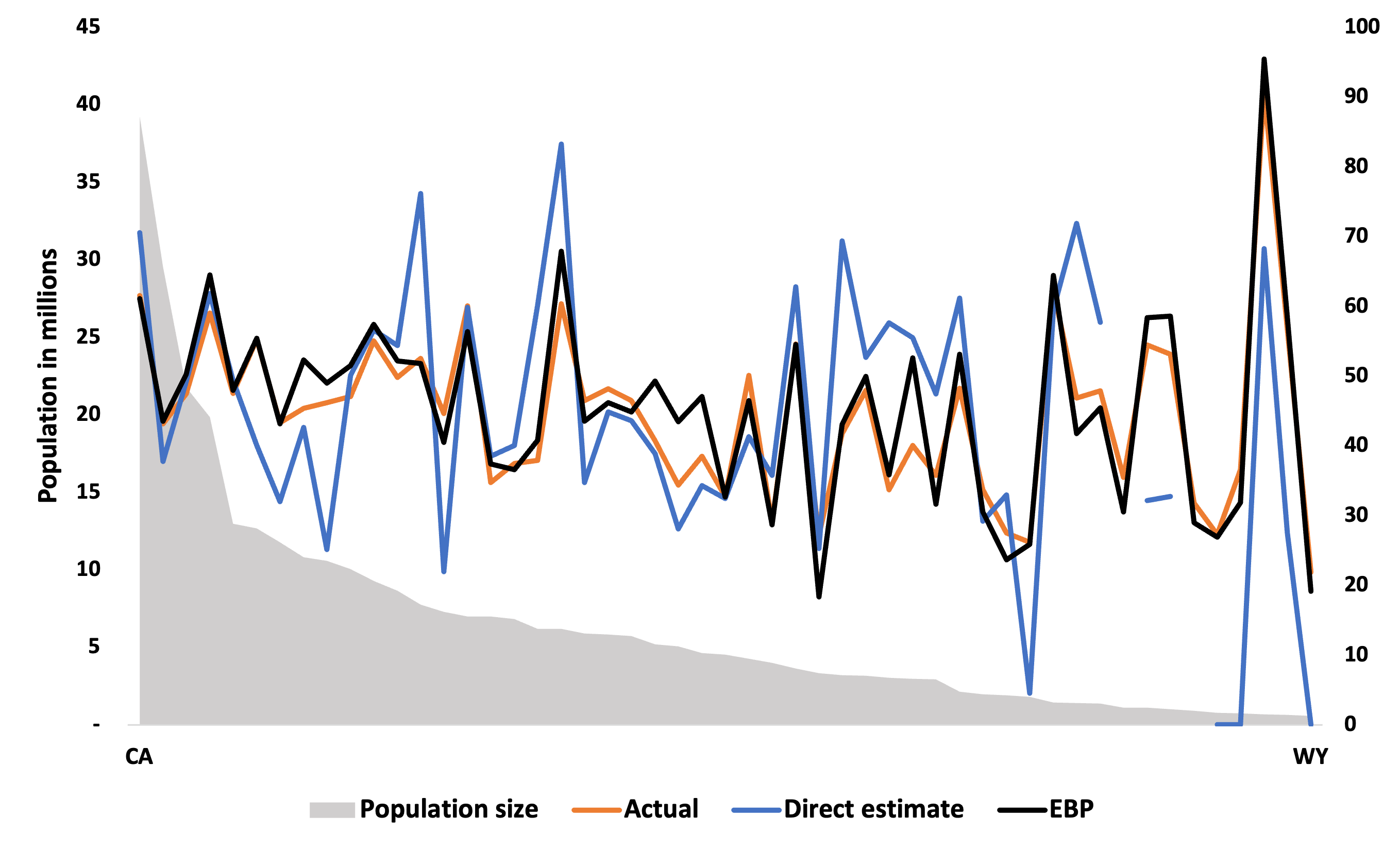}
\end{figure}

\begin{figure}
  \centering
    \begin{tabular}{@{}c@{}}
    \includegraphics[width=.7\linewidth,height=200pt]{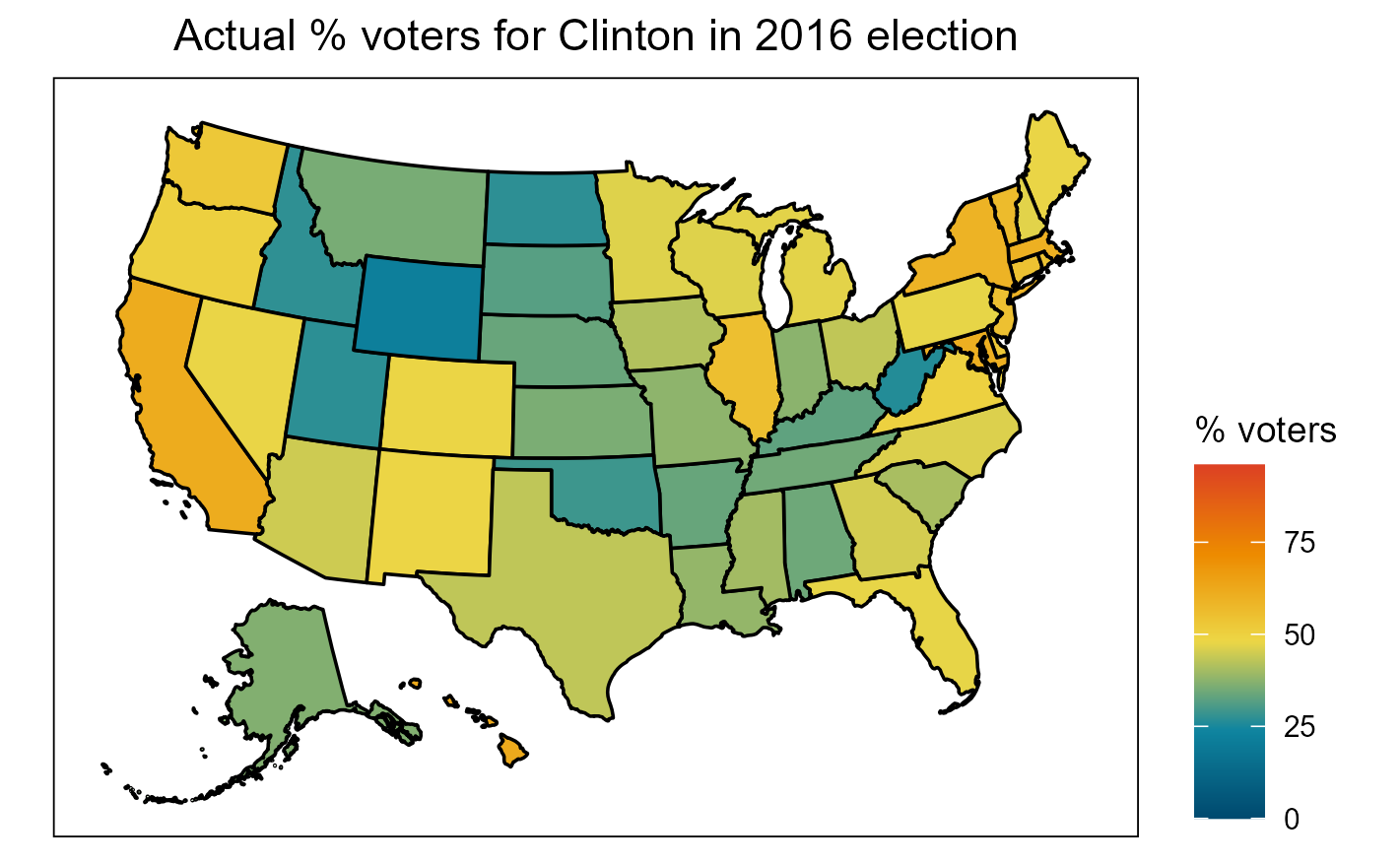} \\[\abovecaptionskip]
  \end{tabular}

  \vspace{\floatsep}

  \begin{tabular}{@{}c@{}}
    \includegraphics[width=0.9\linewidth,height=250pt]{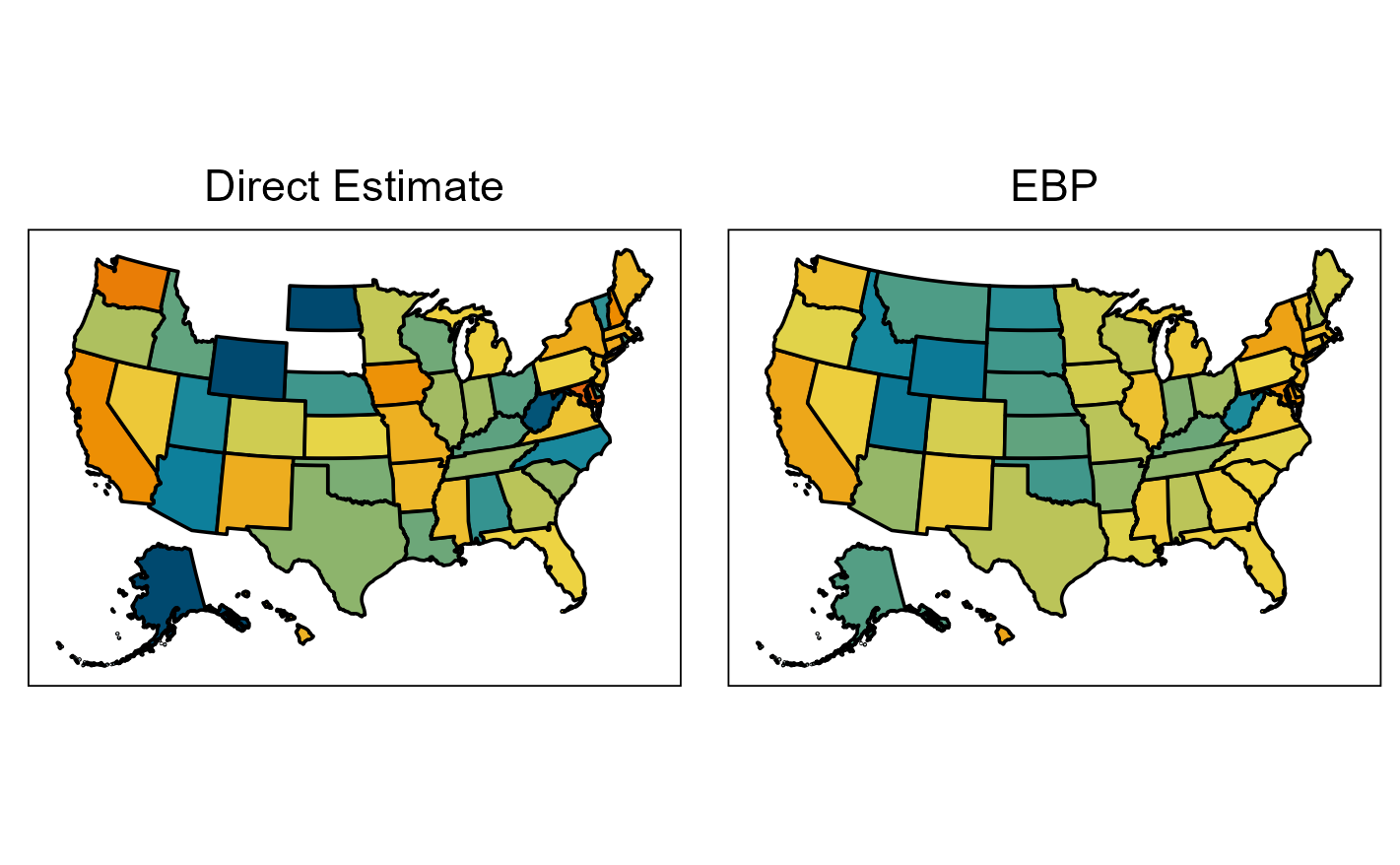} \\[\abovecaptionskip]
  \end{tabular}
\caption{U.S. maps of actual and predicted percentages of voters for Clinton using direct and EBP method.} 
\label{fig:us maps}
\end{figure}

We obtain direct estimates for the 48 matched states via the \texttt{survey} package in R. The estimates and population sizes are noted for a handful of states in Table \ref{tab : state_est}. In general, the direct estimators perform poorly for less populated states with very small samples from the Pew survey. For example, the direct estimates for Arkansas (AK) and Wyoming (WY) are 0 and are entirely missing for MT and SD. However, it is unrealistic that the true percentage of voters for Clinton in any state is zero.  The direct estimate for DC is 68\%, far from the actual percentage (90\%), whereas the EBP estimate for DC (95\%) is very close to the actual. The estimates for all states are displayed in Figure \ref{fig: all_state}. We have ordered the 51 areas in decreasing order of population size for easier comparison. Comparing the actual values with two types of predictors we see that the EBP is working better than the direct method. Compared to the blue line for the direct estimates, the black line for EBPs is much closer to orange line for actual values. The missing direct estimates are noticed in places where the blue line is broken. We also provide similar visualization through maps in Figure \ref{fig:us maps}.

\vskip .2in

\begin{table}[ht]
\caption{Comparison of direct and EBP estimates with actual values denoted in \% for a few states along with their populations sizes. Population sizes in CA, FL, and MD are in millions (denoted by M); the remaining populations sizes are in thousands (denoted by K).} \label{tab : state_est}
\begin{center}
\begin{tabular}{rrrrr}
State & Population size & Actual & Direct & EBP \\
\hline
    CA & 39 M & 61.5 & 70.5 & 61.1\\
    FL & 21 M & 47.4 & 49.2 & 50.2\\
    MD & 6 M & 60.3 & 83.2 & 67.9\\
    MT & 110 K & 35.4 & & 30.5\\
    SD & 895 K & 31.7 & & 28.9\\
    AK & 732 K & 36.6 & 0 & 31.8\\
    DC & 670 K & 90.9 & 68.2 & 95.4\\
    WY & 578 K & 21.9 & 0 & 19.1\\
\end{tabular}
\end{center}
\end{table}

To compare  EBP estimates with direct estimates, we consider three criteria defined in Table \ref{tab: measure}: Average Squared Deviation (ASD), Root Average Squared Deviation (RASD), and Average Absolute Deviation (AAD).  Table \ref{tab: measure} clearly shows that the EBPs  outperform the direct estimates based on all the three criteria.
\begin{table}[ht]
\caption{Comparison of ASD, RASD, and AAD values of EBP and Direct estimator and the respective relative gains (calculated as Direct divided by EBP) at the national level. \label{tab: measure}}
\begin{center}
\begin{tabular}{rrrrr}
Measure & Formula & Direct & EBP & Relative gain\\
\hline
    ASD & $ \frac{1}{51}\sum_{i=1}^{51} (\hat{Y}_i^{est} - \hat{Y}_i^{act})^2$ & 234.2 & 18.9 & 12.4\\
    RASD & $\sqrt{ASD}$ & 15.3 & 4.3 & 3.6\\
    AAD & $ \frac{1}{51}\sum_{i=1}^{51} \left| \hat{Y}_i^{est} - \hat{Y}_i^{act} \right| $ & 11.8 & 3.3 & 3.6\\
\end{tabular}
\end{center}
\end{table}

\subsection{Parametric bootstrap and MSPE calculation}

In this section, we use our proposed parametric bootstrap method on the Pew and CPS data to estimate the MSPE of the EBP, computed for each parameter of interest as the squared difference of the estimated and true values averaged over the $m$ areas. We compare the square root of the estimated MSPE from the bootstrap method with the SE of direct estimates at the state level. Applying the procedure described in section \ref{sec: par boot} to the Pew and CPS data independently to B bootstrap samples, we perform the following steps:
\vskip.2in

\begin{itemize}
\item  [Step 1.] Given $(x_{ij}, \hat{\beta},\hat{\sigma}^2)$ we generate $v_i$ from $N(0,\hat{\sigma}^2)$, compute $\theta_{ij}=\frac{\exp (x'_{ij}\hat{\beta}+v_i)}{1+\exp (x'_{ij}\hat{\beta}+v_i)}$ and generate $y_{ij}$ from $Bern(\theta_{ij}); i = 1,\cdots,m; j = 1,\cdots,c_i$, where $c_i$ can be $\tilde{n}_i$ or $n_i$.
\item  [Step 2.] We generate the $b^{th}$ bootstrap for the Pew data with known auxiliary variables and initial model parameter estimates.
\item  [Step 3.] We apply the algorithm defined in section \ref{sec: EM func} to the Pew bootstrap sample to obtain parameter estimates $(\hat{\beta}_{EM},\hat{\sigma}^2_{EM})$.
\item  [Step 4.] We compute weighted estimates of percentages of voters for Clinton for each of the 50 states and DC using $x'_{ij}$s from the CPS data and Pew parameter estimates obtained in step 3. Hence, $\hat{\bar Y}_{bi}^{EBP}, i=1,\cdots,m$ are the estimated values with respect to the MSPE definition.
\item  [Step 5.] Similarly, generate a CPS bootstrap sample and obtain $\bar y_{biw}, i=1,\cdots,m$ from the $b^{th}$ CPS bootstrap sample. These are the true values with respect to the MSPE definition, i.e., assuming the model is true.
\item [Step 6.] Compute the squared difference between the estimated and true values, i.e., $T_{bi} = \left [ \hat{\bar Y}_{bi}^{EBP} -  \bar y_{biw}   \right ]^2$.
\end{itemize}

Finally, we compute the average over $B$ bootstrap samples to obtain parametric bootstrap MSPE estimate:  $T_i=\frac{1}{B}\sum_{b=1}^B T_{bi}$. To determine the appropriate number of bootstrap samples ($B$), we computed Monte Carlo Error (MCE) for $B=100$ and $500$, where MCE is defined as $\sqrt{\frac{Var(T)}{B}}$. The MCE for $B=500$ is $1.5$, compared to the corresponding value of $3.4$ with  $B=100$. This reduction in error offsets the increase in runtime ($5$ hours for $B=100$, versus more than a day for $B=500$). Table \ref{tab: MSPE and CV} provides the minimum, maximum and average square root MSPE and coefficient of variation (CV) values across $m=51$ areas.


\begin{table}[ht]
\caption{Minimum (Min), Maximum (Max), Average (Avg) of estimated standard errors and coefficients of variations (CV) of EBP for 50 states and DC in percent. The standard error of EBP is the square root of the MSPE, and the CV is the ratio of estimated standard error and estimate. The standard errors are estimated using the proposed parametric bootstrap method using $500$ bootstrap replications.}  \label{tab: MSPE and CV}
\begin{center}
\begin{tabular}{r|rrr|rrr}
 &  & $\sqrt{\widehat{MSPE}}$ & & & CV & \\
   Bootstrap size & Min  & Max & Avg  & Min  & Max & Avg\\ \hline
    100 & 3.5 & 6.9 & 4.9 & 1.1 & 1.7 & 1.4\\
    500 & 2.4 & 6.4 & 4.9 & 1.3 & 1.7 & 1.4\\
\end{tabular}
\end{center}
\end{table}

\begin{figure}[ht]
    \centering
    \caption{Comparison of estimated standard errors of direct estimates and EBPs for the 50 U.S. states and DC. Direct SE is the standard error of the direct estimator obtained from R package \texttt{survey}. The square root MSPE (estimated standard error of the EBP) is estimated using parametric bootstrap method using $500$ bootstrap replications.}
    \includegraphics[width=0.7\linewidth]{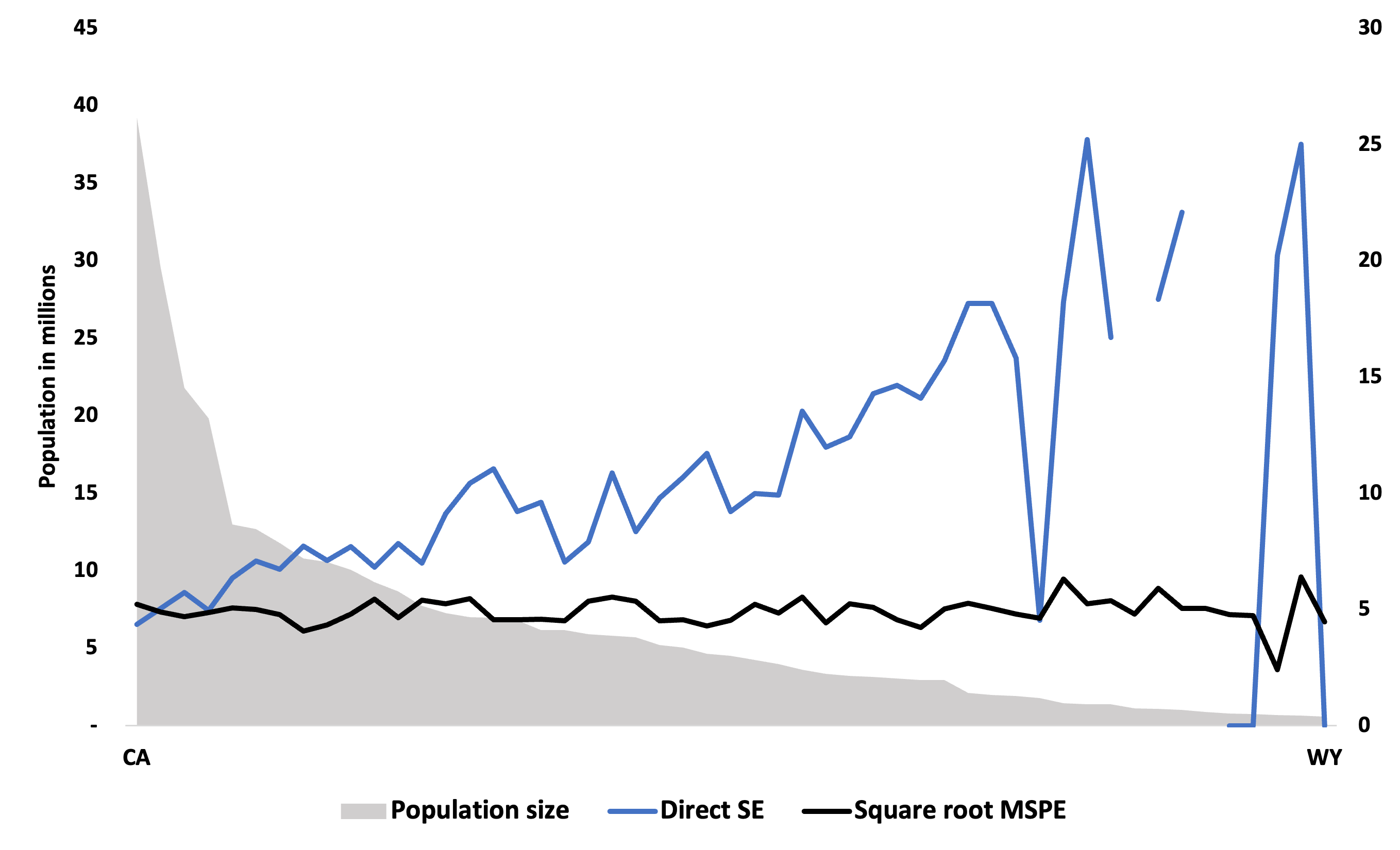}
    \label{fig: se_comparison}
\end{figure}

\noindent Figure \ref{fig: se_comparison} displays the plot of direct estimator SE and EBP SE ($B=500$) for 51 areas. The direct estimates' SE (plotted in blue) varies highly as population size decreases, whereas the EBP square root MSPE (plotted in black) has a more stable trend even as population size decreases with all values in and around 5\%. Table \ref{tab: error_comp} presents direct and EPB standard error estimates for a subset of states. The direct estimates are missing or zero for MT, SD, AK, and WY, in contrast to their reasonable (non-zero) EPB counterparts.

\begin{table}[ht]
\caption{Comparison of estimated standard errors of direct and EBP estimates for selected states (in percent). The standard error of the direct estimator is estimated using R package \texttt{survey}. The standard error of EBP is the square root of MSPE obtained with the proposed parametric bootstrap method using $500$ bootstrap replications.}\label{tab: error_comp}
\begin{center}
\begin{tabular}{rrrrrrr}
 State & SE (Direct) & SE (EBP) \\ \hline
    CA & 4.3 & 5.2\\
     FL & 5.7 & 4.7\\
     MD & 7.0 & 4.5\\
     MT &   & 4.8\\
     SD &   & 5.1\\
     AK & 0 & 4.7\\
     DC & 20.2 & 2.4\\
     WY & 0 & 4.5\\
\end{tabular}
\end{center}
\end{table}

\section{Conclusion}
\label{sec:conc}

In this paper, we develop a new statistical data integration method for estimating small area proportions. 
Unlike standard model-based prediction methods in finite population sampling, we predict the outcome variable for units of a large probability sample survey containing many potential auxiliary variables instead of a sampling frame containing only a limited number of such variables.  Thus, our proposed method can take advantage of a large set of potentially useful auxiliary variables common to both small sample and large probability sample surveys in substantially improving the predictive model.
Our empirical results demonstrate superiority of the proposed empirical best predictors over the usual survey-weighted proportions. In our empirical investigation, we observe that estimates of model parameters could fall on the boundary for some parametric bootstrap samples. The problem is especially prominent for over-fitted models. Thus, we reiterate the importance of model building exercise.  We propose an adjusted maximum likelihood method that ensures parameter estimates to be always within boundaries. 
The proposed data integration method can be used even if the big sample is a nonprobability survey using weights constructed by methods such as the ones described in \citet{chen2020}, \citet{2021valliant} and \citet{savitsky:2022}.  
In this connection, we refer to \citet{nandram2023bayesian} who described a data integration technique using a nonprobability sample and a relatively small probability sample. In their setting both the samples contain the outcome variable $Y$, unlike the situation we propose in this paper where $Y$ is present only in the small probability sample. They used the method of \citet{chen2020} to calculate the weights for the nonprobability sample and propose a Bayesian approach for integrating the two samples whilst we take a frequentist approach.

\bibliographystyle{chicago}
\bibliography{referencesSL.bib}

\begin{thebibliography}{}

\bibitem[\protect\citeauthoryear{Bates, M{\"a}chler, Bolker, and Walker}{Bates et~al.}{2015}]{Bates2015glmer}
Bates, D., M.~M{\"a}chler, B.~Bolker, and S.~Walker (2015).
\newblock Fitting linear mixed-effects models using lme4.
\newblock {\em Journal of Statistical Software\/}~{\em 67\/}(1), 1--48.

\bibitem[\protect\citeauthoryear{Butar}{Butar}{1997}]{Butar1997}
Butar, B. (1997).
\newblock {\em Empirical Bayes methods in survey sampling}.
\newblock Ph{D} thesis, University of Nebraska-Lincoln, Lincoln, Nebraska.

\bibitem[\protect\citeauthoryear{Chatterjee and Lahiri}{Chatterjee and Lahiri}{2007}]{ChatterjeeLahiri2013}
Chatterjee, S. and P.~Lahiri (2007).
\newblock A simple computational method for estimating mean squared prediction error in general small-area model.
\newblock In {\em Proceedings of the Section on Survey Research Methods (JSM 2007)}, pp.\  3486--3493.
\newblock \href{http://www.asasrms.org/Proceedings/y2007/Files/JSM2007-000812.pdf}{http://www.asasrms.org/Proceedings/y2007/Files/JSM2007-000812.pdf}.

\bibitem[\protect\citeauthoryear{Chen, Li, and Wu}{Chen et~al.}{2020}]{chen2020}
Chen, Y., P.~Li, and C.~Wu (2020).
\newblock Doubly robust inference with nonprobability survey samples.
\newblock {\em Journal of the American Statistical Association\/}~{\em 115\/}(532), 2011--2021.

\bibitem[\protect\citeauthoryear{Cochran}{Cochran}{1977}]{Cochran77}
Cochran, W. (1977).
\newblock {\em Sampling Techniques}.
\newblock John Wiley \& Sons, Inc.

\bibitem[\protect\citeauthoryear{Gerber and Furrer}{Gerber and Furrer}{2019}]{GerberOptimParallel}
Gerber, F. and R.~Furrer (2019).
\newblock optimparallel: An {R} package providing a parallel version of the l-bfgs-b optimization method.
\newblock {\em The R Journal Vol. 11/01\/}, 1--6.

\bibitem[\protect\citeauthoryear{Ghosh}{Ghosh}{2020}]{Ghosh2020}
Ghosh, M. (2020).
\newblock Small area estimation: its evolution in five decades.
\newblock {\em Statistics in Transition New Series\/}~{\em 21}, 1--22.

\bibitem[\protect\citeauthoryear{Ghosh and Lahiri}{Ghosh and Lahiri}{1987}]{GhoshLahiri1987}
Ghosh, M. and P.~Lahiri (1987).
\newblock Robust empirical bayes estimation of means from stratified samples.
\newblock {\em Journal of the American Statistical Association\/}~{\em 82}, 1153--1162.

\bibitem[\protect\citeauthoryear{Hall and Maiti}{Hall and Maiti}{2006}]{Hall2006}
Hall, P. and T.~Maiti (2006).
\newblock On parametric bootstrap methods for small area prediction.
\newblock {\em Journal of the Royal Statistical Society: Series B\/}~{\em 68}, 221--238.

\bibitem[\protect\citeauthoryear{Hirose and Lahiri}{Hirose and Lahiri}{2018}]{HiroseLahiri2018}
Hirose, M. and P.~Lahiri (2018).
\newblock { Estimating variance of random effects to solve multiple problems simultaneously}.
\newblock {\em The Annals of Statistics\/}~{\em 46\/}(4), 1721--1741.

\bibitem[\protect\citeauthoryear{Jiang}{Jiang}{2001}]{Jiang2001}
Jiang, J. (2001).
\newblock Mixed-effects models with random cluster sizes.
\newblock {\em Statistics \& Probability Letters\/}~{\em 53}, 201--206.

\bibitem[\protect\citeauthoryear{Jiang}{Jiang}{2007}]{Jiang2007}
Jiang, J. (2007).
\newblock {\em Linear and Generalized Linear Mixed Models and Their Applications}.
\newblock Springer.

\bibitem[\protect\citeauthoryear{Lahiri and Li}{Lahiri and Li}{2009}]{LahiriLi2009}
Lahiri, P. and H.~Li (2009).
\newblock Generalized maximum likelihood method in linear mixed models with an application in small area estimation.
\newblock {\em Proceedings of the Federal Committee on Statistical Methodology Research Conference\/}.
\newblock \href{https://nces.ed.gov/FCSM/pdf/2009FCSM\_Lahiri\_II-C.pdf} {https://nces.ed.gov/FCSM/pdf/2009FCSM\_Lahiri\_II-C.pdf}.

\bibitem[\protect\citeauthoryear{Lahiri and Salvati}{Lahiri and Salvati}{2023}]{Lahiri2023}
Lahiri, P. and N.~Salvati (2023).
\newblock A nested error regression model with high-dimensional parameter for small area estimation.
\newblock {\em Journal of the Royal Statistical Society Series B: Statistical Methodology\/}~{\em 85\/}(2), 212--239.

\bibitem[\protect\citeauthoryear{Li and Lahiri}{Li and Lahiri}{2010}]{LiLahiri2010}
Li, H. and P.~Lahiri (2010).
\newblock Adjusted maximum method for solving small area estimation problems.
\newblock {\em Journal of Multivariate Analysis\/}~{\em 101}, 882--892.

\bibitem[\protect\citeauthoryear{Little and Rubin}{Little and Rubin}{2019}]{LittleRubin2015}
Little, R. and D.~Rubin (2019).
\newblock {\em Statistical analysis with missing data, 3rd edition}.
\newblock Wiley.

\bibitem[\protect\citeauthoryear{Nandram and Rao}{Nandram and Rao}{2024}]{nandram2023bayesian}
Nandram, B. and J.~Rao (2024).
\newblock Bayesian integration for small areas by supplementing a probability sample with a non-probability sample.
\newblock {\em Statistics and Applications {ISSN 2454-7395 (online)}\/}~{\em 22}, 343–374.

\bibitem[\protect\citeauthoryear{Pfeffermann}{Pfeffermann}{2013}]{Pfeffermann2013}
Pfeffermann, D. (2013).
\newblock New important developments in small area estimation.
\newblock {\em Statistical Science\/}~{\em 28}, 40--68.

\bibitem[\protect\citeauthoryear{Pfeffermann and Sverchkov}{Pfeffermann and Sverchkov}{1999}]{pfeffermann1999}
Pfeffermann, D. and M.~Sverchkov (1999).
\newblock Parametric and semi-parametric estimation of regression models fitted to survey data.
\newblock {\em Sankhy{\=a}: The Indian Journal of Statistics, Series B\/}, 166--186.

\bibitem[\protect\citeauthoryear{Rao and Molina}{Rao and Molina}{2015}]{Rao2015}
Rao, J. N.~K. and I.~Molina (2015).
\newblock {\em Small Area Estimation, 2nd Edition}.
\newblock Wiley.

\bibitem[\protect\citeauthoryear{Savitsky, Williams, Gershunskaya, and Beresovsky}{Savitsky et~al.}{2023}]{savitsky:2022}
Savitsky, T., M.~Williams, J.~Gershunskaya, and V.~Beresovsky (2023).
\newblock Methods for combining probability and nonprobability samples under unknown overlaps.
\newblock {\em Statistics in Transition new series\/}~{\em 24}, 1--34.

\bibitem[\protect\citeauthoryear{Seidenberg, Moser, and West}{Seidenberg et~al.}{2023}]{pricssa}
Seidenberg, A.~B., R.~P. Moser, and B.~T. West (2023).
\newblock {Preferred Reporting Items for Complex Sample Survey Analysis (PRICSSA)}.
\newblock {\em Journal of Survey Statistics and Methodology\/}~{\em 11\/}(4), 743--757.

\bibitem[\protect\citeauthoryear{Sugden and Smith}{Sugden and Smith}{1984}]{sudgen1984}
Sugden, R.~A. and T.~M.~F. Smith (1984).
\newblock Ignorable and informative designs in survey sampling inference.
\newblock {\em Biometrika\/}~{\em 71\/}(3), 495--506.

\bibitem[\protect\citeauthoryear{Verret, Rao, and Hidiroglou}{Verret et~al.}{2015}]{verret2015}
Verret, F., J.~N. Rao, and M.~A. Hidiroglou (2015).
\newblock Model-based small area estimation under informative sampling.
\newblock {\em Survey Methodology\/}~{\em 41\/}(2), 333--348.

\bibitem[\protect\citeauthoryear{Wang, Valliant, and Li}{Wang et~al.}{2021}]{2021valliant}
Wang, L., R.~Valliant, and Y.~Li (2021).
\newblock Adjusted logistic propensity weighting methods for population inference using nonprobability volunteer-based epidemiologic cohorts.
\newblock {\em Stat Med.\/}~{\em 40\/}(4), 5237--5250.

\bibitem[\protect\citeauthoryear{Yoshimori and Lahiri}{Yoshimori and Lahiri}{2014}]{YoshimoriLahiri2014}
Yoshimori, M. and P.~Lahiri (2014).
\newblock { A second-order efficient empirical Bayes confidence interval}.
\newblock {\em The Annals of Statistics\/}~{\em 42\/}(4), 1233--1261.

\end{thebibliography}

\end{document}